\documentclass[aps,prl,superscriptaddress,amsfonts,amsmath]{revtex4}
\usepackage{graphicx}
\usepackage{amsmath}
\usepackage{amsfonts}
\usepackage{amssymb}

\newcommand{\Deff}{D_{\text{eff}}}

\begin{document}

\title{Brownian dynamics of a microswimmer}

\author{Vladimir Lobaskin}
\affiliation{Physics Department, Technical University of Munich,
D-85748 Garching, Germany}

\author{Dmitry Lobaskin}
\affiliation{Bear Stearns International, 1 Canada Square, London,
E14 5AD, UK}

\author{Igor Kulic}
\affiliation{School of Engineering and Applied Sciences, Harvard
University, Massachusetts 02138, USA}

\date{\today}

\begin{abstract}
  We report on dynamic properties of a simple model microswimmer composed of three
  spheres and propelling itself in a viscous fluid by spinning motion of the spheres
  under zero net torque constraint. At a fixed temperature and increasing the spinning
  frequency, the swimmer demonstrates a transition from dissipation-dominated to a
  pumping-dominated motion regime characterized by negative effective friction
  coefficient. In the limit of high frequencies, the diffusion of the swimmer can be
  described by a model of an active particle with constant
  velocity.
\end{abstract}


\maketitle

Attention of physics community to the problem of swimming at low
Reynolds numbers, which is relevant for the world of
microorganisms, was attracted by Edward Purcell in seventies
\cite{Purcell}. He formulated the basic principles of
self-propulsion and suggested a variety of simple model
microswimmers that would propel themselves in the Stokesian regime
using non-reciprocal cyclic moves. The most famous of them,
three-link swimmer, was solved analytically only recently
\cite{Stone}. As the Stokesian regime is characterized by the
absence of time in the flow equations, the description of
self-propulsion reduces to a purely geometrical problem of
transformation of the microswimmer's body shape. The problem was
solved for various nearly spherical objects, whose surface is
deformed by a wave-like perturbation in the manner of ciliated
microorganisms
\cite{Stone2,Shapere1,Shapere2,Felderhof1,Felderhof2}. A number of
other simple models performing one- or two-dimensional
non-reciprocal moves as well as their swimming performance were
discussed recently \cite{Najafi,Avron,Felderhof3}.

Recent advances in micromanipulation techniques made it possible
to construct artificial swimmers mimicking the bacterial and
protozoal self-propulsion mechanisms. Most of these machines,
however, are supposed to be driven by an external fields rather
than ATP hydrolysis. The first working device imitating the
flagellum beating and driven by an external magnetic field was
reported recently \cite{Bibette}. Realistic implementations of
DNA-based nanomachines using the ratchet principle were also
suggested \cite{Igor1,Igor2}. Other directions in development of
self-propelling micromachines is related to use of anisotropic
environments, active surfaces or chemical reactions, whose
mechanical response has an inherent asymmetry
\cite{Camalet,Kim,Lammert,Golestanian2,Mano,Paxton}. The focus of
these publications is the propulsion mechanism as such and the
dynamics of a swimmer on long timescales is usually not addressed.
One should note, however, that on the nanometer and micrometer
scale the thermal fluctuations are expected to compete with the
propulsion mechanisms and, therefore, the interplay between the
swimming and dissipation processes is of great interest
\cite{schweitzer,ebeling,Tilch,vicsek,vicsek1,vicsek2}. In this
work, we report on general dynamic properties of a microswimmer at
finite temperatures.

For our study we chose a simple model swimmer, consisting of three
spheres with their centers comprising an equilateral triangle. The
distances between the spheres are fixed. To impose propulsion, we
make the spheres spin via applied constant torque. We arrange the
torque directions as shown in Fig. 1a so that the net torque is
always zero. This constraint mimics the rotation due to internal
degrees of freedom, as it would be in case of a microorganism
swimming. At the same time, the algorithm allows us to keep
control over the amount of supplied energy. The dynamics of the
three-ball animals was modeled numerically using a hybrid
molecular dynamics/ Lattice Boltzmann (LB) simulation method
\cite{ahlrichs,NJP}. The spheres were modeled with the raspberry
setup (a tethered network of point particles wrapped around a big
Lennard-Jones (LJ) bead) \cite{NJP} with the radius $R=3$, while
the links by the massless harmonic bonds of length $L=10$ of high
spring constant between the ball centers. The length units are
defined by the LB cell size. Most simulations were performed in a
cubic box of side $40$ units with periodic boundary conditions.
The dynamic viscosity of the fluid in the LJ units was set to
$\eta=2.55$ if not specified otherwise, and the temperature to
$T=5$. To simulate the action of finite temperature, a fluctuating
stress tensor technique was used and in addition stochastic random
force was applied to each of the beads composing the swimmer
\cite{ahlrichs}. The velocity of each bead $\mathbf{v}$ is then
calculated from the following Langevin equation
\begin{equation}
m\frac{d\mathbf{v}}{dt}=-\zeta(\mathbf{v}-\mathbf{u}(\mathbf{r}))+\mathbf{f}%
(t)+\mathbf{F}_{\mathtt{ext}}+\mathbf{F}_{\mathtt{int}}\label{newton}%
\end{equation}
where $m$ is the bead mass, $\zeta$ the friction coefficient,
$\mathbf{u}(\mathbf{r}))$ the interpolated fluid velocity at the
particle position $\mathbf{r}$, $\mathbf{f}(t)$ is a Gaussian
white noise force with zero mean, whose strength is given via the
standard fluctuation-dissipation theorem in the absence of the
external force, $\mathbf{F}_{\texttt{int}}$ the interaction force
between different beads comprising the raspberry (see Ref.
\cite{NJP} for details), and $\mathbf{F}_{\texttt{ext}}$ the
``external'' force driving the sphere rotation. The magnitude of
the external force was chosen to produce the desired spinning
velocity {\boldmath$\omega$}$= \mathbf{T} /(8 \eta \pi R^3)$,
where the net torque $\mathbf{T} = \sum_i${\boldmath $\rho$}$_i
\cdot \mathbf{F}_i$ is a sum of the moments acting on each surface
bead of the sphere, $\rho$ is the distance from the rotation axis
to the $i$th bead. The energy influx is then given by a sum
$\varepsilon = \sum_i \mathbf{F} \cdot \mathbf{v}_i$ over all
beads constituting the spheres or simply $\varepsilon = \mathbf{T}
\cdot${\boldmath$\omega$}.
\begin{figure}
\begin{center}
\vskip 0.2in
\includegraphics[clip,width=3.7 cm]{pscheme}
\includegraphics[clip,width=4.3 cm]{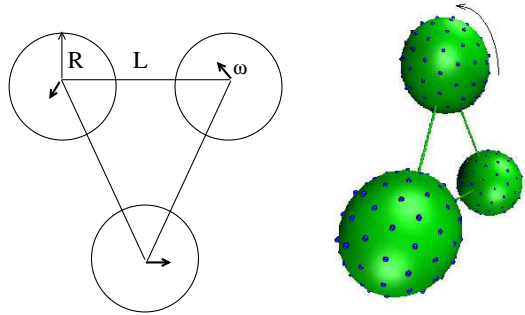}
\end{center}
\caption{
  The construction scheme of the three-ball swimmer: The balls are
  spinning  around the in-plane axes so that the angular velocities
  {\boldmath$\omega$}$_1 +$ {\boldmath$\omega$}$_2 +
  ${\boldmath$\omega$}$_3 = 0$. A simulation snapshot on the right
  hand side illustrates the raspberry model implementation of the
  swimmer.
  }
\label{fig:1}
\end{figure}

\begin{figure}
\begin{center}
\vskip 0.2in
\includegraphics[clip,width=7.5 cm]{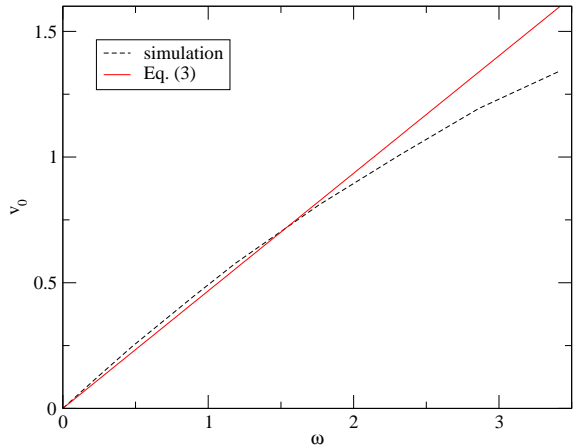}
\end{center}
\caption{
  Self-propulsion velocity of the three-ball swimmer with $R=3$ and $L=10$ as a
  function of the ball spinning frequency. The solid line was obtained using
  Eq. \ref{eq:roll}.}
\label{fig:v}
\end{figure}

Motion of the swimmer at zero temperature can be easily understood
in the limit of small spheres and long links. Each sphere can be
then represented by a rotlet \cite{Landau}
\begin{equation}\label{eq:rotlet}
    \mathbf{v}(\mathbf{r}) =\frac{R^3}{r^3}
    {\texttt\boldmath{\omega}} \times \mathbf{r}
\end{equation}
We have to find the flow field produced by two small spheres at
the location of the third one and take just sum of their the
components normal to the plane containing the three sphere
centers, as the other components cancel. To have a zero net force
acting on each of the spheres, it requires the sphere to move
normally to the triangle plane with the velocity
\begin{equation}
v_0 = \sqrt{3} \omega R \left( \frac{R}{L} \right)^2
\label{eq:roll}
\end{equation}
We tested this prediction against simulation results for a swimmer
with $R=3$ and $L=10$ (see Fig. \ref{fig:v}). Despite the large
ball size as compared to the link length, the equation gives a
very good velocity estimate with relative deviation from the
simulation data less than 4\% up to $\omega=2$. The result given
by Eq. (\ref{eq:roll}) can be also be compared to the propulsion
speed of a twirling ring, which corresponds to a generalization of
the three-ball animal to the case of infinite number of balls
filling a full circle of radius $r$ \cite{Igor1,Igor2}:
\begin{equation}\label{eq:ring}
    v_0 = \frac {R^2}{2 r} \omega \left( \ln{ \left( \frac{8}{R}
\right)} - \frac{1}{2} \right)
\end{equation}

The propulsion efficiency of this swimmer can be estimated as a
relation of the energy dissipated in the forward thrust to the
energy dissipated in the sphere spinning, which gives
\begin{equation}\label{eq:beta}
\beta = \frac{\mathbf{F}_{\texttt{ft}} \mathbf{v}_0}{\mathbf{T}
\cdot{\boldmath \omega}}
\end{equation}
The steady state thrust force ${\mathbf F}_{\texttt{ft}}$ is equal
to the total friction force ${\mathbf F}_{\texttt{f}}$ due to
forward motion of the three-ball swimmer, which is related to the
corresponding mobility as $F_{\texttt{f}} = - v_0 / \mu_3$. If the
hydrodynamic interactions between the spheres are taken into
account in the stokeslet approximation (the mobility of a pair of
stokeslets to the first order in $r^{-1}$ is $\mu_2 =
\frac{\mu_1}{2} (1 + \frac{3}{4} \frac{R}{r})$), we get for the
triplet
\begin{equation}\label{eq:mu} \mu_3 = \frac{\mu_1}{3} \left( 1 + 2
 \frac{3R}{4L} \right) \approx 0.63 \mu_1
\end{equation}
where $\mu_1 = (6 \pi \eta R)^{-1}$ for the motion transversal to
the triangle plane. The torque of the friction force exerted by a
spinning sphere of radius $R$ is given by $T = 8 \pi \eta R^3$. If
we neglect hydrodynamic interactions (due to fast decay of the the
flow field magnitude for a rotlet, $(R/r)^{-3}$, it remains a good
approximation for $(R/L)<1$), the energy dissipated just due to
independent spinning of three spheres becomes $8 \pi \eta R^3
\omega$. The efficiency is then $\beta \approx 0.5 (R/L)^4$, which
reaches its maximum of about 6\% for contacting spheres.

\begin{figure}
\begin{center}
\vskip 0.2in
\includegraphics[clip,width=7.5 cm]{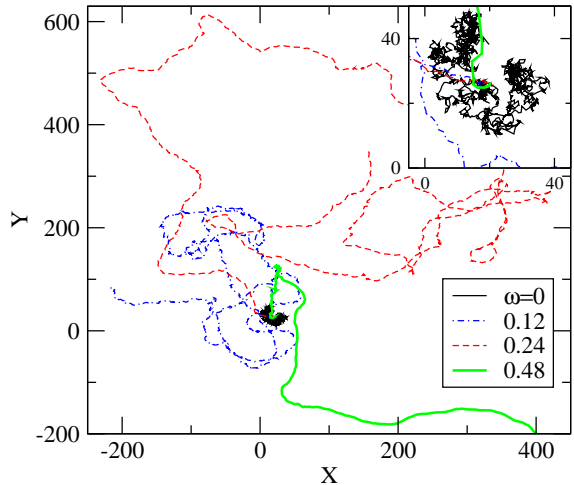}
\end{center}
\caption{
  Two-dimensional XY projection of the trajectories of the swimmer's center of mass
  position obtained with different ball spinning frequencies at
  $T=5$. The insert shows the magnified view of the initial part of the trajectories.
  }
\label{fig:traj}
\end{figure}

We now study the motion of the swimmer at finite temperatures.
Typical two-dimensional projections of the swimmer's trajectories
observed at different ball spinning frequencies are plotted in
Fig. \ref{fig:traj}. One can see that the character of the motion
is changing significantly on increasing the frequency: the
trajectories become more persistent with more extended straight
segments and less abrupt changes of the direction. As a result,
the swimmer with a higher $\omega$ travels over longer distances
within the same time as compared with a less active one.

\begin{figure}
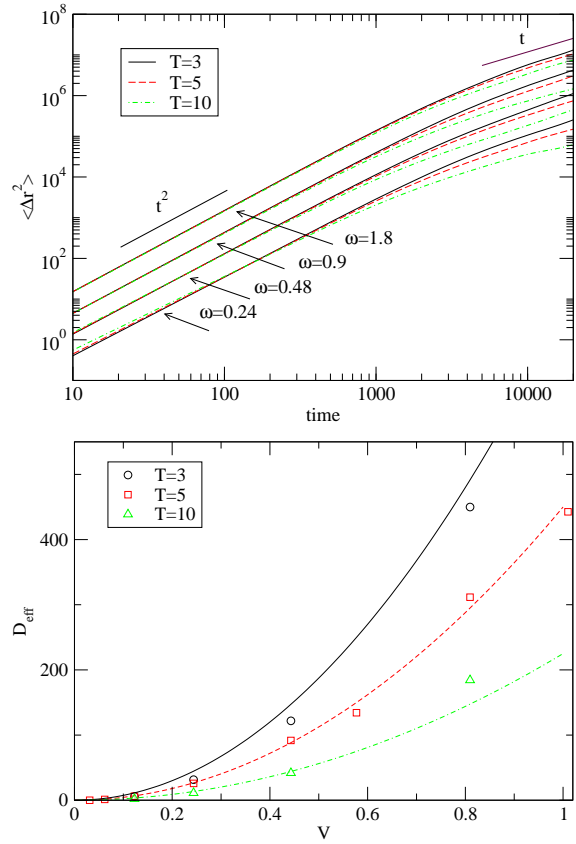

\begin{center}
\includegraphics[clip,width=7.5 cm]{msdall1}
\vskip 0.2cm
\includegraphics[clip,width=7.5 cm]{dpp}
\end{center}
\caption{
  \emph{Top:} Mean-square displacement of the swimmer as a
  function of the ball spinning frequency and temperature. \emph{Bottom:} Long-time
  diffusion coefficient of the swimmer at different temperatures measured from
  the long-time asymptotic behavior of the mean-square displacement. The solid
  lines are calculated as $\Deff=2 v_0 / D_r$  (see the text for further details).
  }
\label{fig:pmsd}
\end{figure}

The mean-square displacements (MSD) of the swimmer's center of
mass are shown in Fig. \ref{fig:pmsd}a for four different spinning
frequencies and three different temperatures. We see immediately,
that the higher spinning frequencies correspond to a higher MSD,
i.e. to the higher diffusivity. The shape of the MSD curves
clearly indicates the existence of two distinct dynamic regimes:
the driven motion at $t \leq 500$ where $\langle \Delta r^2
\rangle \propto t^2$ and the diffusive motion at $t > 1000$ where
$\langle \Delta r^2 \rangle \propto t$. In the case of high
propulsion speeds, the transition from the deterministic driven
motion to a random walk happens due to rotational diffusion of the
swimmer. The ``ballistic part'' of the MSD coincides for the
swimmers with the same frequency but different temperatures. In
contrast, the random walk part is thermosensitive: higher
temperature leads to a lower diffusivity. The transition between
the two regimes happens on the characteristic timescale of
rotational diffusion. The diffusion coefficient measured from the
asymptotic behaviour of the swimmer at long times, $\Deff$, is
plotted in Fig. \ref{fig:pmsd}b. We see that the measured
diffusion coefficient grows nonlinearly with the drift velocity
(or the spinning frequency as $v_0 \propto \omega$).

\begin{figure}
\begin{center}
\vskip 0.2in
\includegraphics[clip,width=7.5 cm]{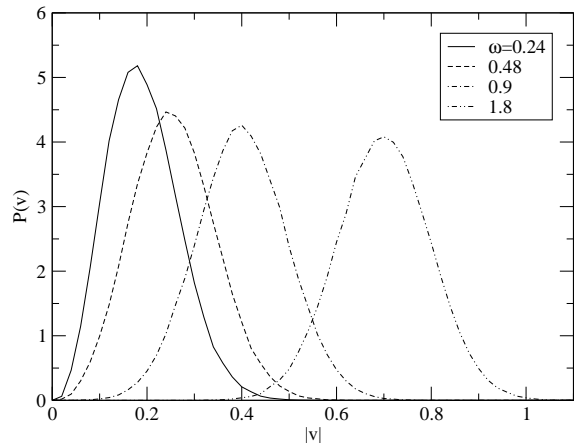}
\end{center}
\caption{
  Distribution function of the swimmer's velocity
  component in the direction of the thrust force at different
  spinning frequencies and temperature $T=3$.
  }
\label{fig:pv}
\end{figure}

A simplest model describing the diffusion of an active swimmer
moving with a constant velocity $v_0$, whose direction of motion
changes due to rotational diffusion with a diffusion constant
$D_r$, gives the diffusion coefficient $D_{\texttt{eff}} = 2
v_0^2/D_r$. This expression follows immediately from a
reformulation of the known result for the end-to-end distance of a
persistent polymer chain, which is $\langle \Delta r^2 \rangle = 2
l_p L$, $l_p$ being the persistence length and $L$ the contour
length of the chain. In the swimmer variables, the persistence
length of the trajectory is $l_p=\tau v_0$, where $\tau$ is the
rotational correlation time, and its contour length is simply
proportional to time: $L= t v_0$. Then, $D_{\texttt{eff}} =
\langle \Delta r^2 \rangle / 6t = v_0^2 \tau /3$ \cite{Lovely}.
The time $\tau$ is defined via the orientational correlation
function of the swimmer $\langle \mathbf{n}(0) \cdot \mathbf{n}(t)
\rangle$, where $\mathbf{n}$ is the normal vector to the plane of
the triangle connecting the sphere centers. In case of random
uncorrelated fluctuations, the correlation function decays at
short times exponentially with time as $\langle \mathbf{n}(0)
\cdot \mathbf{n}(t) \rangle = e^{-t/\tau}$. In case of an
anisotropic rotation, one can expect that the decay is governed by
the smallest correlation time, i.e. by the characteristic time of
rotation in the direction with the smallest friction torque. The
shortest orientational relaxation time expected for this swimmer
corresponds to a rotation around the in-plane axis passing through
the center of one ball and through the middle of the spring
connecting the opposite ones. With a first order correction for
hydrodynamic interaction between the balls we get
\begin{equation}\label{eq:tau}
\tau = \frac{1}{k_B T} \left (8 \pi \eta R^3 + \frac{3 \pi \eta
L^2 R}{1 - 3R/2L} \right)
\end{equation}
This formula gives $\tau = 2250$, $1350$, and $675$ for $T=3$,
$5$, and $10$, respectively. This values are close to what we
observe in simulation at low $\omega$. If we use this correlation
time to predict the translational self-diffusion coefficient, the
agreement with the measured data is convincing (the solid curves
in Fig. \ref{fig:pmsd}b).
We also tested our assumption regarding the orientational
correlation time by direct measurement of the correlation function
of the swimmer trajectories. From these measurements at low
spinning frequencies we get $\tau = 2300 \pm 200$,  $1300 \pm 200$
and $700 \pm 100$ for $T=3$, 5, and 10, respectively. These values
are in good agreement with the estimated correlation times. In
what concerns the assumption of constant velocity along the
trajectory, we can check the swimmer's velocity distribution. The
characteristic distributions of the velocity component
perpendicular to the triangle plane (i.e. parallel to the thrust
force) for $T=3$ are shown in Fig.\ref{fig:pv}. While at
$\omega=0.24$ the variance is comparable to the mean velocity, at
the highest frequency, $\omega=1.2$, it does not exceed 25\% of
the mean. So, we see that the swimmer's velocity is narrowly
distributed around $v_0$ and that also this assumption is
justified. Thus, the swimmer's motion, especially at the high ball
spinning frequencies (or energy influx rates) fits well the model
of an active Brownian particle \cite{schweitzer,ebeling}, whose
velocity is fairly constant in the magnitude while the direction
is subject to thermal fluctuations. In other properties we observe
more analogies to the active Brownian particle model. For example,
the existence of a finite preferential velocity in our model is
also reflected in the velocity distribution function shown in Fig.
\ref{fig:pv}. Instead of a maximum at $v = 0$ characteristic for
systems in thermal equilibrium (the Maxwell distribution) it
possesses two maxima at $\pm v_p$ and a minimum at $v=0$.

\begin{figure}
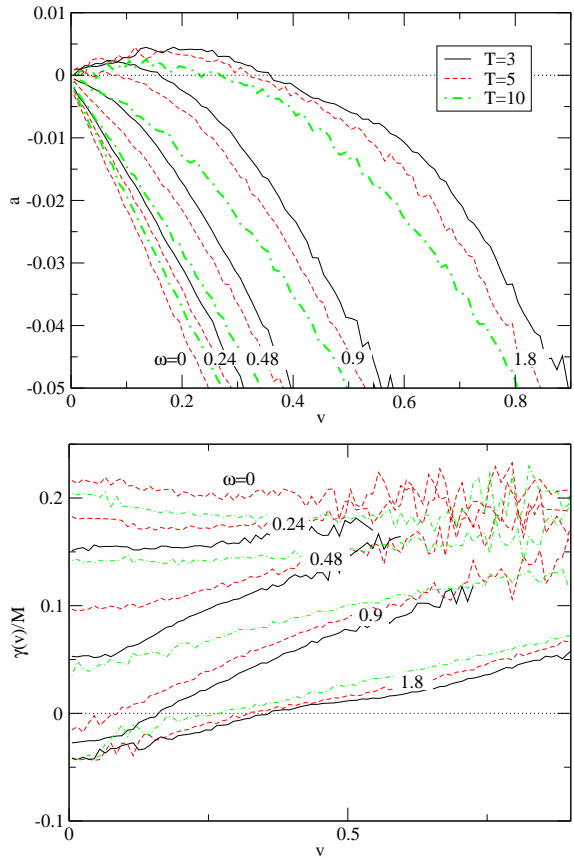

\begin{center}
\vskip 0.2in
\includegraphics[clip,width=7.5 cm]{v-a310}
\vskip 0.2cm
\includegraphics[clip,width=7.5 cm]{f310}
\end{center}
\caption{
  \emph{Top:} Swimmer acceleration as a function of its instant velocity at different
  ball spinning frequencies (marked on the curves) and temperatures.
  \emph{Bottom:} The corresponding effective friction coefficient as a function of
  the swimmer's velocity. }
\label{fig:pva}
\end{figure}

At $v_0^2 > \langle v^2(\omega = 0) \rangle$ we expect that the
character of the swimmer's motion changes. As it was mentioned
above, the characteristic velocity $v_0$ is a function of the
energy influx rate, while the variance is fixed by the temperature
and the swimmer mass: $\langle v^2 \rangle = 3 k_B T / M$, and
characterizes the energy dissipation rate. Thus, the transition is
expected at $\omega = k_BTL^4/(R^6 M)$. The threshold values of
the frequency are $\omega \approx 0.3$, 0.5 and 1.0 (or
$v_0=0.13$, 0.23, and 0.45) for $T=3$, $5$ and 10, respectively.
At the higher frequencies, other signatures of the driven regime
can be observed, such as negative friction coefficient
\cite{ebeling}. The friction coefficient can be extracted from the
relation between the swimmer acceleration $\mathbf{a}$ and its
instant velocity $\mathbf{v}$ . We calculated these values by
averaging over three velocity/acceleration components. The result
for different spinning frequencies is shown in Fig. \ref{fig:pva},
top. As there is no other external forces acting of the swimmer,
the acceleration is resulting from the viscous friction $a=
F_{\mathtt{fr}}/M = - \gamma(v) v/M$. Here, the effective friction
coefficient $\gamma(v)$ is a function of the instant velocity. The
$a(v)$ curves for the lowest spinning frequency are fairly linear
at low velocities, which corresponds to a constant positive
friction coefficient. The curves bend toward smaller friction at
the higher frequencies, and at the highest one all the curves
cross zero and display a region of positive acceleration. All the
$a(v)$ curves show two linear regions: one at $v \to 0$ and
another one at the higher velocities. The corresponding nonlinear
friction coefficient is plotted in Fig. \ref{fig:pva}, bottom.
There we indeed notice saturation both at low velocities and at
high $v$. While the limiting values $\gamma (v \to 0)$ shows
strong dependence on the energy influx rate, the high velocity
values seem to tend to a temperature independent limit, which only
slightly decreases with the ball spinning frequency. We also
notice that the parts of the curves with the acceleration parallel
to the instant velocity correspond to $\gamma(v)<0$. One can see
that the transition into the driven regime with negative $\gamma$
occurs on increasing $\omega$. At $T=3$ and $T=5$ the negative
friction appears at $\omega=0.9$ and 1.2, at $T=10$ only at
$\omega=1.2$, in agreement with the above estimates. The values
$v_p$, at which the friction coefficient turns zero reflect
therefore the preferential stationary velocity, at which the
viscous resistance of the medium is exactly balanced by the
swimmer's forward thrust.

We would like to note also that the behaviour of the friction
coefficient qualitatively resembles the velocity dependence
predicted within the active Brownian particle model
\cite{ebeling}. In the adiabatic approximation the effective
friction for a particle with an internal energy depot should vary
as $\gamma (v) = \gamma_0 - \frac{a}{b + v^2}$, where the
coefficient $a$ is proportional to the energy influx rate. If
applied to our case, the $\gamma_0$ would be the Stokesian viscous
friction experienced by the swimmer. The negative values of the
nonlinear friction coefficient can be observed provided that
$\gamma (v)$ turns negative at finite velocities, which requires
that $a > \gamma_0 (b + v^2)$. We can see from Fig.\ref{fig:pva}
at $0.24 \leq \omega \leq 0.9$ that the shape of the friction
curves $\gamma(v)$ shows a similar behaviour.

Finally, we would like to note that the parameters of this system
were chosen to demonstrate the wide range of dynamic properties.
In case of a micrometer size living swimmers, the role of thermal
fluctuations might be significantly smaller. Living microswimmers
moving at the velocities of 10$\mu$m/s would rather demonstrate
the driven motion regime. Still, in presence of a stochasticity
caused by internal mechanisms of the animal or spatial/temporal
inhomogeneities of the medium, its competition with the
deterministic drive would lead to qualitatively similar results.
On the other hand, slow and small swimmers like the artificial
ones, suggested in literature, would demonstrate many features
described in this work even with thermal fluctuations. Another
parameter that differentiates the dynamics studied in the present
work from the biological microorganisms is the relatively high
Reynolds number that reaches $\texttt{Re}_{max} \approx 3$ for the
highest velocities considered. So, the role of inertial effects is
considerably overestimated as compared to the microworld. Our
tests, however, showed that none of the reported properties can be
attributed solely to inertia and we expect our predictions to be
valid also for the typical microscopic objects.

We studied dynamics properties of a self-propelling microswimmer
engine subjected to thermal fluctuations. Our model for the first
time treats explicitly both the propulsion mechanics and
fluctuations and demonstrates a rich interplay of these factors.
Qualitatively, many of the observed properties (such as the
non-linear velocity-dependent friction coefficient, characteristic
crater-like velocity distributions at high energy influx rates,
etc.) resemble those of the model of active Brownian particle
\cite{schweitzer,ebeling}. The differences related to the
fluctuations of the propulsive effort, the broad velocity
distributions, three-dimensional motion, and anisotropic friction,
result in general in a more complex dynamic behaviour, which we
hope will stimulate further theoretical effort in this field.

\begin{acknowledgements} We are grateful R.
Thaokar, W. Ebeling, L. Schimansky-Geier, U. Erdmann, B. U.
Felderhof, J.-F. Joanny for stimulating discussions.
\end{acknowledgements}

\end{document}